# A Large Area Detector proposed for the Large Observatory for X-ray Timing (LOFT)

S. Zane*[a], D. Walton[a], T. Kennedy[a], M. Feroci[b], J.-W. Den Herder[c], M. Ahangarianabhari[d], A. Argan[e], P. Azzarello[f], G. Baldazzi[g], D. Barret[h], G. Bertuccio[d], P. Bodin[i], E. Bozzo[f], F. Cadoux[l], P. Cais[m], R. Campana[b], J. Coker[a], A. Cros[h], E. Del Monte[b], A. De Rosa[b], S. Di Cosimo[b], I. Donnarumma[b], Y. Evangelista[b], Y. Favre[l], C. Feldman[n], G. Fraser[n], F. Fuschino[o], M. Grassi[p], M.R. Hailey[a], R. Hudec[q], C. Labanti[o], D. Macera[d], P. Malcovati[p], M. Marisaldi[o], A. Martindale[n], T. Mineo[r], F. Muleri[b], M. Nowak[s], M. Orlandini[o], L. Pacciani[b], E. Perinati[t], V. Petracek[u], M. Pohl[l], A. Rachevski[v], P. Smith[a], A. Santangelo[t], J.-Y. Seyler[i], C. Schmid[w], P. Soffitta[b], S. Suchy[t], C. Tenzer[t], P. Uttley[z], A. Vacchi[v], G. Zampa[v], N. Zampa[v], J. Wilms[w], B. Winter[a]
on Behalf of the LOFT Consortium

[a]Mullard Space Science Laboratory, UCL, Holmbury St Mary, Dorking, Surrey, RH56NT,UK, [b]INAF-IAPS-Roma via Fosso del Cavaliere, 100, 00133, Rome, Italy, [c]SRON, The Netherlands Institute of Space Research, Utrecht, The Netherlands, [d]Politecnico di Milano, Como campus, via Anzani 42, 22100, Como, Italy, [e]INAF HQ, Viale del Parco Mellini 84, 00136, Roma, Italy, [f]ISDC, Science Data Center for Astrophysics, Ch. D'Ecogia 16, 1290, Versoix, Geneva, Switzerland, [g]University of Bologna, Dept. of Physics and INFN section of Bologna, V.le Berti Pichat, 6/2, 40127, Bologna, Italy, [h]Institut de Recherche en Astrophysique et Planetologie, IRAP, 9 Avenue du Colonel Roche, BP44346, 31028, Toulouse, France, [i]Centre National d'Etudes Spatiales, Centre Spatial de Toulouse, 18 Avenue Edouard Belin, 31 401, Toulouse, CEDEX 9, France, [l]DPNC, Geneva University, Quai Ernest-Ansermet 24, CH-1211,Geneva, Switzerland, [m]Laboratoire d'Astrophysique de Bordeaux, Univ. Bordeaux, CNRS, UMR5804, BP 89, 33270 Floirac – France, [n]Space Research Centre, Department of Physics and Astronomy, University of Leicester, Leicester, LE17RH, UK, [o]INAF/IASF Bologna, via Gobetti 101, 40129, Bologna, Italy, [p]University of Pavia, via Ferrata 1, 27100, Pavia, Italy, [q]Czech Technical University in Prague, Faculty of electrical Engineering and Astronomical Institute, Academy of Science of the Czech Republic, Ondrejov, Czech Republic, [r]INAF/IASF Palermo, via Ugo la Malfa 153, 90146, palermo, Italy, [s]MIT, NE80-6077, 77 Massachusetts Ave., Cambridge, MA 02139, [t]IAAT, University of Tuebingen, Sand 1, 72076, Tuebingen, Germany, [u]Czech Technical University in Prague, Faculty of Nuclear Science, Prague, Czech Republic, [v]Istituto Nazionale di Fisica Nucleare, INFN, Sezione di Trieste, Padriciano 99, I-34149, Trieste, Italy, [w]Dr Remeis-Observatory & ECAP, University of Erlangen-Nuremberg, Sternwartstr. 7, 96049 Bamberg, Germany, [z]Astronomical Institute Anton Pannokoek, University of Amsterdam, Postbus 94249, 1090 GE Amsterdam.

**ABSTRACT**

The Large Observatory for X-ray Timing (LOFT) is one of the four candidate ESA M3 missions considered for launch in the 2022 timeframe. It is specifically designed to perform fast X-ray timing and probe the status of the matter near black holes and neutron stars. The LOFT scientific payload is composed of a Large Area Detector (LAD) and a Wide Field Monitor (WFM). The LAD is a 10 $m^2$-class pointed instrument with 20 times the collecting area of the best past timing missions (such as RXTE) over the 2-30 keV range, which holds the capability to revolutionize studies of X-ray variability down to the millisecond time scales. Its ground-breaking characteristic is a low mass per unit surface, enabling an effective area of ~10 $m^2$ (@10 keV) at a reasonable weight. The development of such large but light experiment, with low mass and power per unit area, is now made possible by the recent advancements in the field of large-area silicon detectors - able to time tag an X-ray photon with an accuracy <10 µs and an energy resolution of ~260 eV at 6 keV - and capillary-plate X-ray collimators. In this paper, we will summarize the characteristics of the LAD instrument and give an overview of its capabilities.

## 1. INTRODUCTION

High time resolution X-ray observations of compact objects provide direct access to strong field gravity, black hole masses and spins, and the equation of state of ultradense matter, hence inputs to particle physics not testable in the laboratory and unique tests of general relativity.

LOFT (the Large Observatory for X-ray Timing) is one of the four candidate ESA M3 missions, proposed for a launch in the 2022 timeframe[1]. LOFT is specifically designed to answer the ESA's question "matter under extreme conditions" by exploiting the diagnostics of very rapid X-ray flux and spectral variability that directly probe the motion of matter down to distances close to black holes horizon and neutron stars surface. The LOFT payload comprises the Large Area Detector (LAD), a 10 $m^2$-class instrument with 20 times the collecting area of the best past timing missions (such as RXTE, the largest predecessor), which holds the capability to revolutionize the studies of X-ray variability from X-ray sources on the millisecond time scales. The LAD will operate in the energy range 2-30 keV (up to 80 keV in expanded mode) with good spectral resolution (<260 eV @ 6 keV Full Width Half Maximum, FWHM) and a temporal resolution of 10 µs.

The key to the 20x breakthrough in effective area achieved by the LAD resides in the synergy between technologies imported from other fields of scientific research, both ground- and space-based. In particular, the crucial ingredients for a sensitive but lightweight experiment, enabling ~15 $m^2$ geometric area payload at reasonable weight, are the innovative large-area Silicon Drift Detectors (SDDs) designed on the heritage of the ALICE experiment at CERN/LHC[2,3]. These will be combined with a collimator based on lead-glass micro-capillary plates (the mechanical structure of the well-known microchannel plates).

The drift concept makes the spectroscopic performance of the SDDs weakly dependent on the extent of the collecting surface: large-area (~75 $cm^2$) monolithic detectors can be designed, with only 224 read-out anodes (thus low power requirements, ~20 W $m^{-2}$) but still very good spectral performance. This design allows to achieve, for the first time, an unprecedented large throughput (~2.5 x $10^5$ cts $s^{-1}$ from the Crab) with a segmented detector, making pile-up and dead-time, often worrying or limiting focused experiments, secondary issues.

The timescales that LOFT will investigate range from sub-millisecond quasi-periodic oscillations (QPOs) to months long transient outbursts, and the relevant objects include many that flare up and change state unpredictably, so relatively long observations, flexible scheduling and continuous monitoring of the X-ray sky are essential for the success of the mission.
For this reason, LOFT will also be equipped with a coded-mask Wide Field Monitor (WFM, 2-50 keV), using the same type of detectors as the LAD (see a separate contribution in this volume[4]). The WFM will provide coverage of about 66% of the sky every pointing (largely exceeding the field of view of any past X-ray monitor) therefore discovering new transients and spectral changes, triggering pointed observations on the most interesting and extreme states with either the LAD or other multi-wavelength facilities in synergy with the radio and optical surveys set to start in the near future.

*sz@mssl.ucl.ac.uk; www.ucl.ac.uk/mssl

## 2. SCIENCE OBJECTIVES

The main goal of the LOFT mission is the study of neutron stars and black holes, i.e. of the configurations which possess the strongest gravitational fields in the universe. They provide unique opportunities to reveal for the first time a variety of general relativistic effects and thereby investigate gravity in the strong-field regime, and to measure fundamental parameters of collapsed objects gaining important insights into the physics of matter at supranuclear densities and in supercritical magnetic fields. In particular, the LOFT top level goals encompass:

- the study of ultradense matter and neutron star structure, via accurate measurements of spin pulsations, neutron star masses and radii, and through the astroseismological study of crustal oscillations following intense flares from Soft Gamma Repeaters. Understanding the properties of ultradense matter and determining its equation of state (EOS) is one of the most challenging problems in contemporary physics. At densities exceeding that of atomic nuclei, exotic states of matter such as Bose condensates or hyperons may appear; and at even higher densities a phase transition to strange quark matter may take place. Only neutron stars have the potentiality to probe these densities in the "zero" temperature regime relevant to these transitions.
- the study of strong field gravity close to black holes and neutron stars, via measurements of black holes mass and spins from time variability and spectroscopy, general relativistic precession, epicyclic motion, quasi-periodic oscillations (QPOs) evolution and Fe line reverberation studies in bright AGNs.

In order to achieve the top level goals, the effective area of the LAD is crucial. Only such a large instrument can lead, for instance, to measure the mass and radius of a neutron star with an accuracy better than 5% (a powerful probe for the EOS of the internal matter), to measure global seismic oscillations for faintest flares (opening a new window on astroseismology) or to follow QPO's evolution in the time domain (which will unveil the fundamental frequencies of the motion of matter in the inner, strong-field gravity-dominated regions of the black holes disks). LAD's much improved energy resolution (better than 260 eV) compared to that of RXTE/PCA will allow the simultaneous exploitation of spectral diagnostics, in particular the study of the relativistically broadened 6-7 keV Fe-K lines detected in the spectra of Active Galactic Nuclei (AGN) and in turn will lead to measurements of the black hole spin and mass.

With its unprecedented capabilities, the LAD will also be a powerful tool for studying the X-ray variability and spectra of a wide range of objects, from accreting pulsars and bursters, to magnetar candidates (Anomalous X-ray Pulsars and Soft Gamma Repeaters), cataclysmic variables, bright AGNs, X-ray transients and the early afterglows of Gamma Ray Bursts. Through these studies it will be possible to address a variety of problems in the physics of these objects. Coordinated optical/NIR and radio campaigns on specific themes, as well as spin measurements which can aid the Advanced Virgo/LIGO searches for gravitational wave signals from fast rotating neutron stars will add great value to the program and will make the LAD a favorite instrument for a community even larger than that traditionally involved in X-ray timing.

## 3. LOFT-LAD REQUIREMENTS AND DESCRIPTION OF THE MAIN INSTRUMENT COMPONENTS

The LAD payload is a large array of X-ray detectors with a total geometric area of 15 $m^2$. The current baseline for the configuration is based on 6 detector panels (approx 1x3 $m^2$ each), connected by hinges to an optical bench at the top of a tower, in an arrangement similar to that previously used for SAR antenna wings, with the WFM on the top of the tower (see Figure 1). The satellite can be stowed into the launcher volume (the original baseline design was based on Vega, while more recent analysis indicate that a Soyuz launch choice would be more feasible), launched into a Low Equatorial Orbit (~600 km) in order to reduce the background and radiation damage from the South Atlantic Anomaly, and deployed in space by means of a deployment mechanism similar to that used in SAR missions.

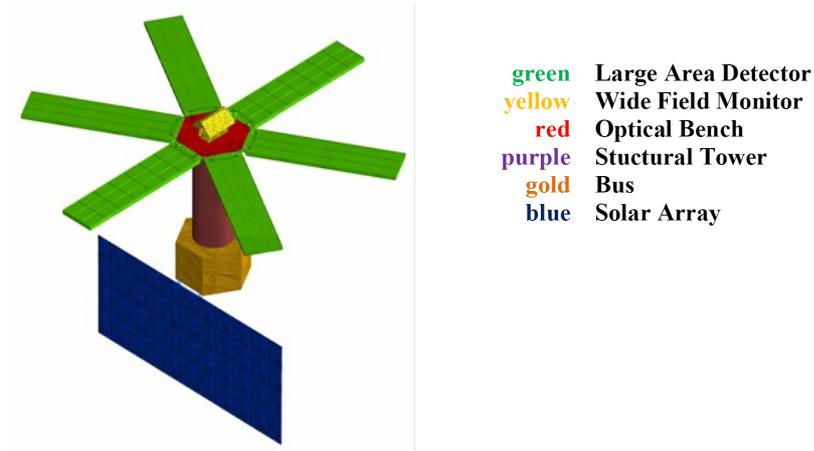

Figure 1. A schematic view of the LOFT spacecraft, showing in colour codes the different elements (the WFM design is in evolution, see a separate paper in this volume for all details[4]).

Table 1 summarizes the LAD scientific requirements, as derived by the science objectives. The study of the energy-resolved timing properties of the X-ray emission of cosmic sources requires the accurate measurement of the time-of-arrival (TOA) and energy of the largest number of photons from the target source. The unambiguous identification of the target source in this type of experiment (e.g., the PCA onboard RXTE[5]) is most effectively achieved by narrowing the field of view by means of an aperture collimator, down to a level (typically ≤ 1°) large enough to allow for pointing uncertainties yet small enough to reduce the aperture background (cosmic diffuse X-ray background) and the risk of source confusion.

In this type of instrument, the knowledge of the impact point of the photon on the detector array is not needed, so there is no need for position sensitive detectors. Instead, detector read-out segmentation is useful/necessary to reduce the effects of pile-up and dead time.

The LAD is therefore designed as a classical collimated experiment. The 6 Detector Panels will be tiled with 2016 SDDs, electrically and mechanically organized in groups of 16, referred to as Modules. Each of the 6 Panels hosts 21 Modules, each one in turn composed of 16 SDDs (see Figure 2).

The SDDs are 450 µm thick and operate in the energy range 2-80 keV. The field of view of the LAD will be limited to <1 degree by X-ray collimators. These are developed by using the technique of micro-capillary plates, the same used for the microchannel plates: a ~6 mm thick sheet of Lead glass is perforated by a huge number of micro-pores, ~100µm diameter, ~ 20 µm wall thickness. The stopping power of Pb in the glass over the large number of walls that off-axis photons need to cross is effective in collimating X-rays below 30 keV.

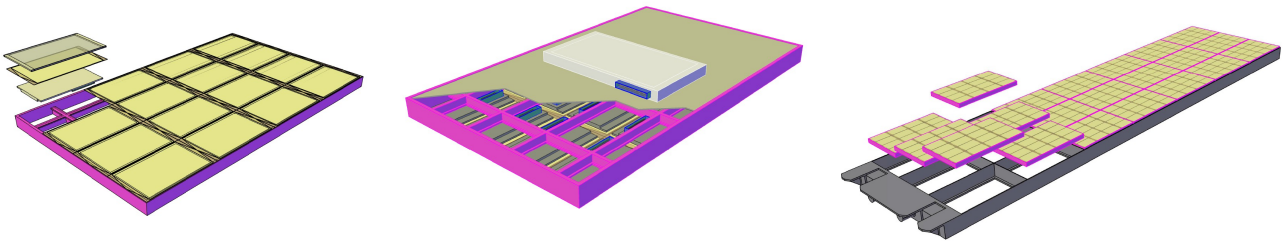

Figure 2. Left: Front-side view of a Module, showing the mounting of the collimator, SDD and the Front End Electronics. Center: Back-side view of a Module, showing the radiative surface and the Module Back End Electronic. Right: a LOFT Detector Panel with all the assembled Modules and the interfaces to the deployment system.

Table 1. Summary of LAD requirements

| Item | Requirement | Goal |
|---|---|---|
| Effective Area | 4m$^2$ @ 2keV | 5m$^2$ @ 2keV |
| | 8m$^2$ @ 5keV | 9.6m$^2$ @ 5keV |
| | 10m$^2$ @ 8keV | 12m$^2$ @ 8keV |
| | 1m$^2$ @ 30keV | 1.2m$^2$ @ 30keV |
| Energy range | 2-80 keV | 1.5-80 keV |
| Energy resolution (beginning of life) | 260 eV @ 6keV | 200 eV @ 6keV |
| | 200 eV (singles, 45%) | 160 eV (singles, 45%) |
| Collimated FOV (FWHM) | < 1 degree | < 0.5 degree |
| Time Resolution | 10μ s | 7μ s |
| Absolute time accuracy | 1μ s | 1μ s |
| Dead time | < 1% at 1 Crab | < 0.5% at 1 Crab |
| Background | <10 mCrab | < 5 mCrab |
| Background knowledge | 1% at 5-10keV | Same |
| Max flux (continuous, rebinned in energy >30keV) | >500 mCrab | >750 mCrab |
| Max Flux (continuous, re-binned) | 15 Crab | 30 Crab |

### 3.1 The Silicon Drift Detectors

The primary enabling technology for the LAD is the large-area SDDs. These were originally developed for the Inner Tracking System (ITS) in the ALICE experiment of the Large Hadron Collider (LHC) at CERN, by one of the scientific institutes in the LOFT Consortium, the INFN Trieste, Italy, in co-operation with Canberra Inc.[2,3,6].

The key properties of the SDDs are their capability to read-out a large photon collecting area with a small set of low-capacitance (thus low-noise) anodes and their very small weight (~ 1 kg m$^{-2}$)[7]. The working principle is shown in Figure 3: the cloud of electrons generated by the interaction of an X-ray photon is drifted towards the read-out anodes, driven by a constant electric field sustained by a progressively decreasing negative voltage applied to a series of cathodes, down to the anodes at ~0 V. The diffusion in Si causes the electron cloud to expand by a factor depending on the square root of the drift time. The charge distribution over the collecting anodes then depends on the absorption point in the detector.

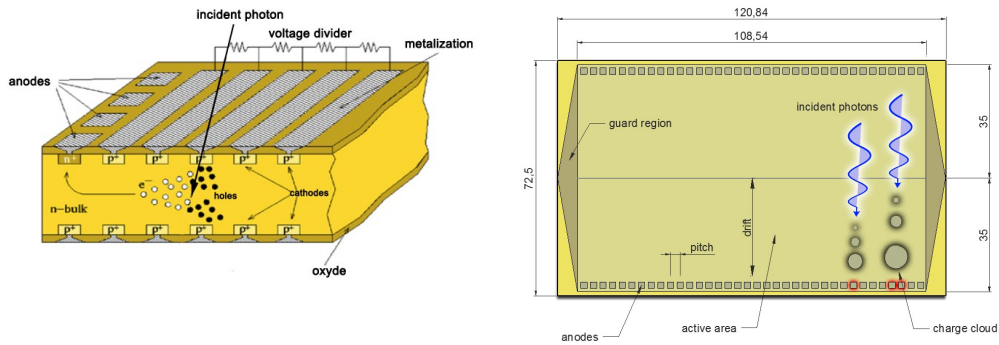

Figure 3. The electrical structure and working principle (left) together with size and functional drawing (right) of a SDD.

The ALICE SDDs have successfully operated since 2008. They were produced on 5-inch diameter, 300 µm thick Si wafers and have a monolithic active area of 53 cm$^2$ each, with an anode pitch of 294 µm. The LAD detector design is an optimisation of the ALICE detector: 6-inch, 450 µm thick wafers will be used to produce 76 cm$^2$ monolithic SDDs (Figure 3: 108.52 mm x 70.00 mm active area, 120.84 mm x 72.50 mm geometric area). The anode pitch will be increased to 970 µm to lower the read-out channel count, thus reducing the power consumption (corresponding to an elemental area of 0.970 mm x 35 mm = 0.3395 cm$^2$).

The Si tile is electrically divided in two halves, with 2 series of 112 read-out anodes at two edges and the highest voltage along its symmetry axis. The drift length is 35 mm. A drift field of 370 V/cm (1300 V maximum voltage) gives a drift velocity of ~ 5 mm/µs and a maximum drift time of ~ 7 µs, as measured at room temperature (+20 °C). This is the highest detector contribution to the uncertainty in the determination of the absolute time of arrival of the photon, still a factor 30% smaller than the scientific requirement. The maximum size of the charge cloud reaching the anodes (depending mostly on the drift distance, not on the photon energy) is ~ 1 mm (corresponding to an event absorbed at the bottom of the drift channel). We note that, depending on the relative size and position of the Gaussian-shaped charge cloud when reaching the anode pattern, the event charge may be collected by 1 or 2 anodes. Based on this, we define the event *multiplicity* as single (approximately 45% of the total), when the full charge of the event is collected by a single anode, and double (~ 55% of the total) when the charge is shared on two neighbouring anodes. Since in the two options the same charge compares to the noise of one or two anodes, the single events display higher spectroscopic quality and therefore can be selected for observations requiring higher spectral performance (hence the two different energy resolution values in Table 1). They correspond to about half of the LAD area. In general, single events correspond to photons absorbed at a relatively small distance from the collecting anodes and in a drift position relatively "centered" with the anode (a photon detected even very close to the anode but in a position that is in between the two drift channels will share its charge over the two anodes anyway).

The large-area SDDs were originally designed for particle tracking, i.e. high energy events. However, over the last few years, work has been carried out to characterise and optimise the same detector design for detection of soft X-rays. The preliminary results obtained with a spare detector of ALICE (300 µm thick, 294 µm anode pitch, no design optimization), with a bread-board read-out based on discrete electronics, show high spectral performance already at room temperature, as shown in Figure 4. The very preliminary results obtained with this prototype in terms of X-ray spectral and position resolution have now been published[8,9].

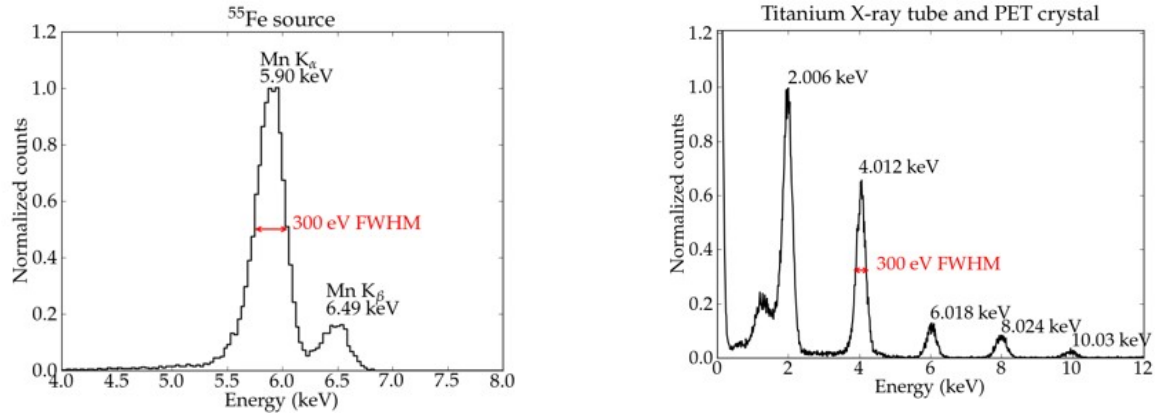

Figure 4. Energy spectra as measured using a spare ALICE detector, equipped with discrete read-out electronics, at room temperature. The measured FWHM is ~300 eV at 5.9 keV. The minimum line energy is ~1.5 keV (corresponding to the spurious Al k-fluorescence from the detector box).

Even if the anode capacitance of the SDD is very small (~ 50 fF), the measurements suffered from a high stray capacitance (due to the fanout scheme required by the discrete read-out electronics) and a large leakage current due to the room temperature (measured as ~ 8 pA, for a volume of 294 µm x 300 µm x 35 mm). The results obtained in the lab at room temperature can be largely improved by using an integrated read-out electronics which allows to optimize the input layout, by lowering the operating temperature and achieving much smaller values of the leakage current, enabling much higher spectral resolution. However, when operated in space, Silicon detectors suffer from a number of effects due to the radiation environment (Total Dose, Non-Ionizing Energy Losses, …) usually causing a severe increase in the leakage current. It has been already understood that, in the LOFT orbit, the radiation environment is very favourable. However, despite the very low particle fluence, NIEL events are still expected to cause a severe increase in the detector leakage current. The amount of the increase strongly depends on the parameters of the orbit (mostly inclination and altitude), as the particles causing the effect are the low energy protons trapped in the South Atlantic Anomaly (SAA). In particular, our analysis shows that operating the detectors at temperature of -30 °C meets the scientific requirements on spectral resolution end-of-life and in the worst case for the orbit selection (orbit at 600 km and ~5° inclination). An inclination of 2.5° would allow to meet the requirements already at -20 °C (or to achieve better spectral performance at -30 °C).

The LAD detection element is undergoing a significant technology development program by the LOFT groups in Italy (INFN, INAF, Polytechnic of Milan and Universities of Pavia and Bologna). LOFT-specific SDDs are being produced in tight collaboration with the Bruno Kessler Foundation (FBK, Trento): already two detector prototype runs have been produced and tested, optimizing the original ALICE design for the low energy response, low-power and anode pitch requirements of the LAD. A third prototype, most representative of the current LAD baseline, has just been submitted for production. Test results with discrete read-out showed performance in line with requirements. In order to reach a full demonstrator of the in-flight configuration, a read-out 32-channel test-ASIC based on the STARX-32 design (Bastia et al. 2010, Proc. AMICSA Workshop) has been developed. The preliminary bench tests of the ASIC prototype showed an ENC of ~ 14 $e^-$ (room temperature), required to be able to test the ultimate SDD spectroscopy performance. A front-end breadboard was designed and produced at DPNC (Geneve). Functional and performance tests are planned starting in Summer 2012. Figure 5 shows the testing system: FEE board (left panel), hosting one detector and 4 ASICs (pictures of real detector and ASIC are shown as in-sets); full set-up box (right panel), including a cooling system for a more representative test of the in-flight operation, allowing interface with the test equipment and the X-ray facility.

In addition to the full-demonstrator activity, an intense program of radiation tests of the SDD prototypes is being carried out by the LAD Team. An ALICE spare detector was used to measure the NIEL radiation damage from 50 MeV protons at the PIF/PSI facility in Villigen, while one of the FBK prototype was used to test specific radiation damage by soft (200 and 800 keV) protons at the accelerator in Rosenau/Tuebingen. A test with debris is also planned within July 2012 at the dust accelerator at MPIK (Heidelberg).

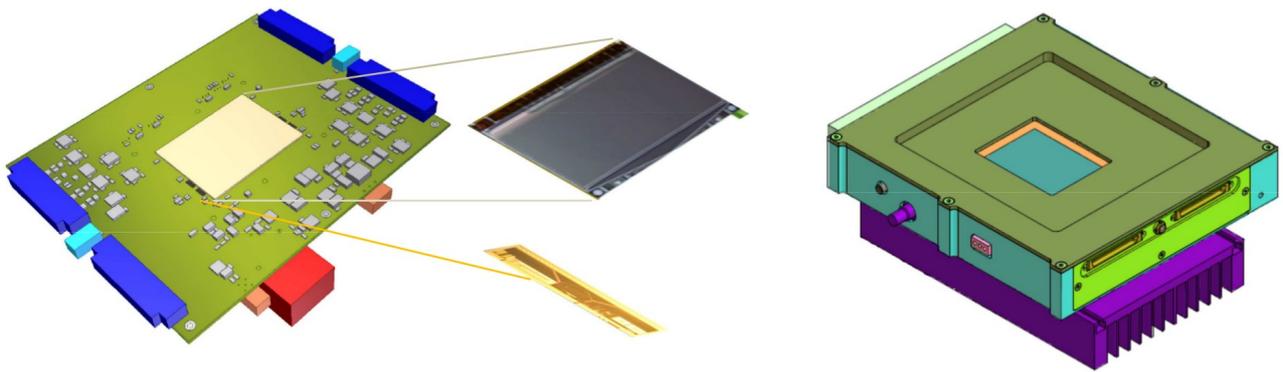

Figure 5: The breadboard front-end electronics board and housing designed and built for the full-demonstrator of the LAD detector (see text for details). Photos of the detector and test-ASIC prototypes are shown.

### 3.2 The Collimator

The other key element of the LOFT payload innovative design is the capillary-plate X-ray collimator. This is based on the technology of microchannel plates (MCPs) and made of a multi-pore, ~ mm thin sheet of lead-glass which will be able to absorb soft X-rays coming from outside its aperture holes.

The MCP drawing is currently studied at SRC (Leicester, UK) based on the heritage of EXOSAT (1983-6), whose MEDA and GSPC detectors were collimated in this way, and on the much more recent development of microchannel plate X-ray optics for the BepiColombo Mercury Imaging X-ray Spectrometer (MIXS) experiment[10]. In particular, the collimator channel of this instrument (MIXS-C) provides an attractive basis for the LAD collimator design. The channel size is about ~20-30 µm, while the walls are as thin as ~4-6 µm. The aspect ratio of the holes ranges typically between ~40 and ~200 (thickness ~few mm) and the open area ratio ranges from 50% to 70%. A key property of these collimators is their very low mass: the net mass per unit area is about ~3 kg m$^{-2}$, enabling large areas at reasonable weight.

The collimator would provide effective shielding from photons arriving from outside the field of view above the primary energy range of LOFT (2-30 keV). Photons with high energy passing through will be detected with low efficiency by the thin Si detector and will be anyway discriminated by their energy deposition in the detector, except for some Compton interactions that will contribute to the residual instrumental background. The same property (i.e. transparency at high energy) will constitute an advantage for different science cases and will be exploited for the detection of bright and hard events (e.g., GRBs, magnetar flares, …) from outside the FoV.

The MIXS-C MCP collimator design, manufactured by Photonis (Brive, France) is highly developed and is already near to the LOFT requirement. The main characteristics of the MIXS-C heritage collimator (in the context of the LOFT application) are the following:
- highly developed (Technology Readiness Level 5)
- square pore sizes 20 µm (Figure 6)
- standard size of individual plate: 40 mm x 40 mm
- open area ratio (OAR) 65%-70%
- orthogonality of channels to 1 arcmin
- Pb content 37%
- 80 nm Aluminium filming of channels provides thermal control surface
- thermal, mechanical and integration issues known
- 4 x 4 plates unit vibration demonstrated
- MCP detector typically sustain ~ 1.0-1.3 kV/mm (e.g., ROSAT or Chandra)

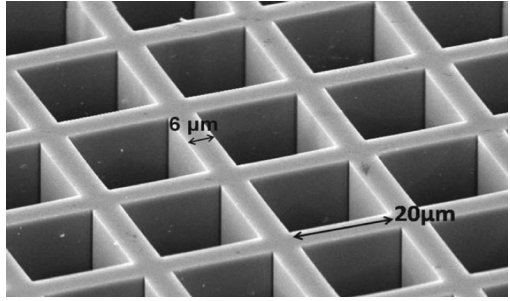

Figure 6. Scanned electron microscope image of a sample MCP collimator.

The MCP optic production process, furthermore, has been demonstrated to be flexible with regard to channel pore size (10-100 µm square) and channel aspect ratio (10:1 - 500:1), which are important aspects that will drive the finalization of the design for LOFT-LAD.

As compared to the LOFT proposal baseline, the MIXS-C heritage collimator has a number of advantages and similarities. On the other hand, it also showed a lower Pb content, a smaller OAR and a smaller plate size.

The originally-proposed LAD collimator parameters have evolved during the study phase to date, and the LAD team has assessed some of the most important requirements, including (i) the field-of-view (ii) the required transparency of the collimator at high (i.e. 30-80 keV) X-ray energies and (iii) the requirement for a "flat-top" component of the basic triangular collimator response function, describing X-ray transmission versus off-axis angle. Accordingly, new key technical findings are now informing the new collimator baseline design. We found that there is no requirement to develop a new microchannel plate glass with higher lead oxide fraction (~37%) than the present standard lead silicate glass, since the stopping power of the collimator can be simply controlled by changing the channel septal thickness. Also, there is no requirement to use radioisotope-free glass in the collimator manufacture, since the count rate due to $^{40}$K betas and gammas is estimated to be well below other background sources.

LOFT is aimed at the observation of point-like sources. Under this perspective, the collimator field of view should be as small as possible, to minimize the number of photons from the isotropic diffuse X-ray background, limiting the instrument sensitivity. The minimum collimator field of view and angular response then derive from the scientific requirement of not introducing detectable spurious timing features by modulating the observed count rate from a point-like source because of a variable instrument response. In practice, this implies having a "stable response" over an angular range large enough to accommodate the internal misalignment of the instrument (permanent and thermo-elastic, including the panel deployment system), as well as the attitude instability. This issue was studied in great detail by the team, detailing and motivating the scientific requirement. The result is a set of requirements on the internal LAD alignment as well as on the spacecraft attitude control system able to guarantee that spurious modulations are always well below any detectable astrophysical signal, at the different frequencies.

As for the plate size, the LAD design is compatible with a conservatively assumption that we could cover each LAD SDD with 6 collimator plates, with a mechanical assembly very similar to MIXS-C and with even smaller plate sizes (40 mm x 40 mm). However, manufacture of larger size plates is feasible and more efficient mountings with larger plate size have now been studied by the team. The current goal for the LAD collimator element is to make use of a self supporting single tile, 11 cm x 8 cm size. An 80 nm, self standing aluminum film will be placed on the MCP front side (including holes), at production level. In addition to the MSSL heritage in MCP charged particle detectors, Leicester has been using similar large structure in spring-mounted photon counting detectors -up to 10 x 10 cm$^2$ and indeed 16 x 5 cm$^2$.

The key elements of the design for the LOFT-LAD collimator are summarized in Table 2. Figure 6 shows a scanning electron microscope image of a MIXS-C format plate; the side length of the channels is 20 µm, the wall thickness is 6 µm.

Table 2. Summary of the LAD collimator design

| Parameter | Value |
| --- | --- |
| Pore size (µm) | 100 |
| Septal thickness (µm) | 20 |
| Channel aspect ratio | 60:1 |
| Channel length | 6 mm |
| Open area fraction | 70% |
| Spherical slumping | None |
| Channel coating | None |
| Leakage | Acceptable to 30 keV |

### 3.3 The ASIC

Under joint management by CNES and IRAP, the LOFT LAD ASIC will be developed under an industrial contract with the DOLPHIN company. The technology retained for the ASIC is the TSMC (Taiwan Semiconductor Manufacturing Company) 0.18 micron. The first design will include the analog part of the ASIC: the charge preamplifier, the shaping amplifier and the peak and hold. The ASIC will include 16 channels, with a pad pitch of around 900 microns. The ASIC will be produced and tested by the end of 2012 to demonstrate that the noise performance (20 e- RMS) and the low power consumption requirements can be achieved. The second run will merge the 16 analog channels with a 11 bit ADC. Fully representative of the flight model, it will be tested by the end of 2013, to demonstrate that simultaneously, all the functionalities, the noise level and the low power consumption (less than 0.650 mW/channel) requirements are met.

## 4. INSTRUMENT CONFIGURATION AND MECHANICAL DESIGN

In the current design, the LAD experiment is composed of 6 independent and identical detector panels. Such a design satisfies the scientific requirements of LOFT in terms of effective area within the envelope of a Vega launcher, that was considered as a baseline during the initial study of the mission. The baseline configuration is not a constraint at present, as alternative configurations that satisfy the same requirements and optimize the overall resources can certainly be considered, especially because the choice of a Soyuz launcher offers additional space (e.g. for panel size).

The basic LAD detection element is composed of SDD+FEE+Collimator, hereafter referred to as Detector. The assembly philosophy employs a hierarchical approach: Detector, Module, Detector Panel, LAD Assembly. The LAD Assembly is composed of 6 Detector Panels (DPs), one DP is composed of 21 Modules. Each Module includes 16 Detectors.

The read-out electronics is organized as follows. Each Detector is equipped with its own Front-End Electronics (FEE). The FEEs of the 16 Detectors in a Module converge into a single Module Back End Electronics (MBEE). One Panel Back-End Electronics (PBEE) for each DP is in charge of interfacing in parallel the 21 MBEE included in a PBEE, making the Module the basic redundant unit. A block diagram of the LAD organization is shown in Figure 7. The digital data processing is described in details in a separate paper in this volume[11].

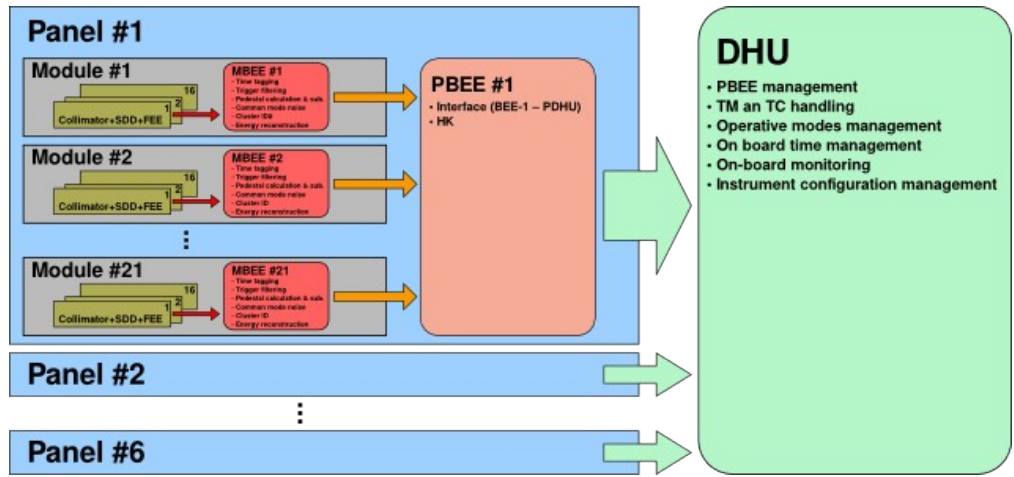

Figure 7. A block diagram showing the organization and structure of the LAD.

The current design envisages a set of 6 co-aligned DPs connected by hinges to a satellite structure, the optical bench hosting the WFM, or the spacecraft bus itself. The panels are folded down during launch and deployed when in orbit. Shielding against X-ray background will be provided by the collimator (front) and a metal shield (back). A further thermal screen filter will contribute (together with the MCP collimator optical properties) to a high level of rejection of UV/Visible/IR but will be transparent to low energy (2 keV) X-rays. For the LAD, each filter can be mounted above each collimator, either at tile level or at detector size. The soft X-ray transparency is >90% at 2 keV and the optical transmission is <$10^{-6}$.

In the following we will briefly describe the currently assumed LAD mechanical design, made of 6 identical detector panels.

### 4.1 Detector Panel

The detector panel is the main structure of the LAD and will hold the 21 detector modules. It will be connected via hinges to the deployment tower, and, thanks to its stiffness, will provide alignment and stability for all the modules. Via its structure the various harnesses will run toward each module.

A set of 21 modules will be integrated into a DP (see right panel in Figure 2), together with the Panel Back-End Electronics (PBEE) box and all the electrical connections and routing between the 21 MBEEs and the single PBEE in each DP. In order to fulfil the alignment requirements, the mechanical interface of the Modules with the DP structure will be of isostatic type. The mechanical frame constituting the DP will be built in Carbon Fibre Reinforced Plastic. The stiffness of the structure and the low coefficient of thermal expansion of this material will in fact reduce significantly thermo-mechanical stresses and thermo-elastic deformations of the DP system.

### 4.2 Module

A design option for a module, containing 16 SDDs, is shown in Figure 8. In order to minimise alignment errors for the micropore optics all optics for the module are held in one large frame; shown in the example as 2 MCP tiles per SDD within the module. This should be regarded as a worst case for the configuration: as mentioned before, the current goal is to make use of a single tile (it is important to emphasise here that the "optical" element for the LAD is the collimator and not the detector, which is only required to collect all the photons transmitted by the collimator).

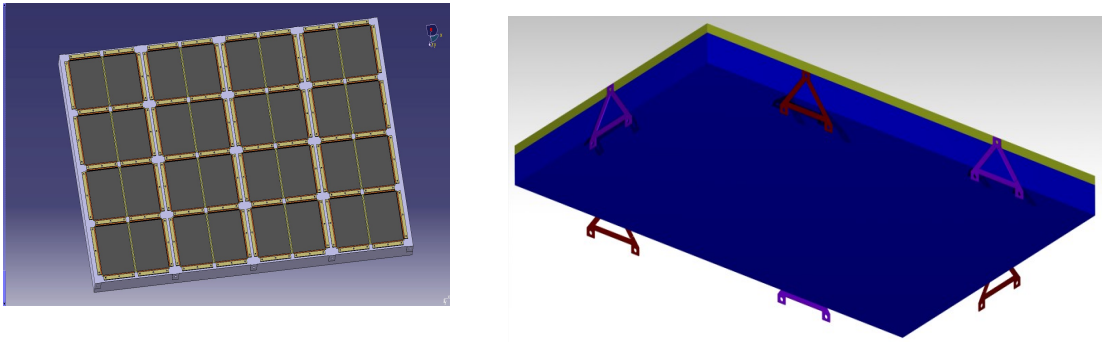

Figure 8. Detector module (left) and its bottom view (right). The clamped collimator is shown here as a two tile layout (worst case for the configuration as far as the clamping is concerned). The Al alloy frame is shown in light grey, the clamping springs in orange, the clamping plates in gold. Note that the design is in evolution ond only three of the isostatic interfaces are now baselined.

The module consists of an aluminium box, (shown in blue, Figure 8 right) holding a printed circuit board which contains all 16 SDDs in a 4 by 4 grid. On the top of the box the collimator tiles are clamped in an aluminium alloy frame.
One advantage of a single tile or two tile collimator is that the tiles can be clamped rather than bonded to the frame, which allows the frame to be made from aluminium alloy, reducing mass and simplifying manufacture and alignment.

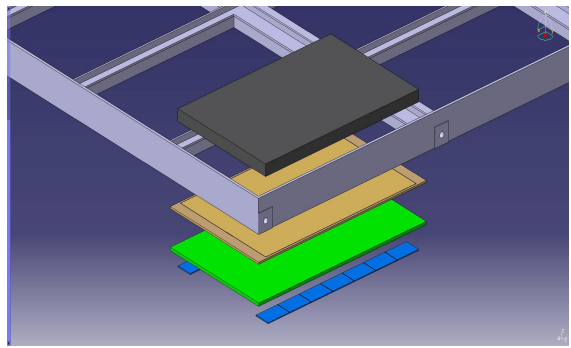

Figure 9. Exploded views of the detector module, single tile detector assembly. In dark grey we see the micropore optic, in light grey is the Al alloy frame, light brown is the SDD tile, green is the FEE-PCD and blue is the ASIC.

In our design, the array of micropore optics tiles are located at the top of the module. They are all set within a Al alloy frame, located just underneath the tiles. The SDD are mounted on a single printed circuit board with the ASICs attached on the other side of the board. The board sits in an aluminium alloy box. In Figure 9 we see an exploded view (single micropore tile concept), with the tile at the top, then the light grey aluminium alloy frame, the SDD (brown), the FEE PCB (green) and the ASICs (blue).

A recent study of the LAD background components has shown that a high level of control of the LAD background systematic can be achieved if a detector surface equivalent to 1 module is equipped with a "blocked collimator". This will be a Pb glass tile, made with the same material and the same stopping power (mass and g cm$^{-2}$) as the collimator, which will allow to monitor the non-aperture background (>90% of the LAD background) continuously. As a baseline the 16 blocked detectors will be placed in a single module, becoming a "blocked module". Future simulations will clarify whether this is an optimal solution, or the blocked detectors are better distributed over different Detector Panels, but always adding up to 16 detectors. The thermo-mechanical properties of these blocked collimators will be nearly identical to those of the real collimators.

## 4.3 Detector

As already mentioned, we refer as "detector" for the assembly of SDD, FEE and collimator. In fact, each Silicon Drift Detector will be equipped with its own read-out electronics. The SDD will be back-illuminated, in order to minimize the electrical contacts on the X-ray entrance window. On the front side of the SDD the Si tile will be glued to the printed circuit board hosting the front-end electronics (ASICs and front-end components). The size of the circuit board is slightly smaller than the size of the Si tile, so that the input pads of the ASICs, hosted at the edge of the board, directly face the anode pads of the SDD, minimizing the length of the wire bonding. The high voltage connection is also on the same (front) side. Instead, the medium voltage (powering the last section of the drift field and the pull-up cathodes) needs to be brought to the X-ray entrance side. The dimensions of each Si tile are 72.5 mm x 120.84 mm (including an active area of 108.54 mm x 70.0 mm), with a thickness of 0.45 mm. The FEE and its integration concept is shown in Figure 10. The FEE board is slightly smaller than the SDD tile to favour the wire-bonding connection to the SDD anode pads. Its dimensions are 66.0 mm x 120.84 mm, with a 2 mm thickness.

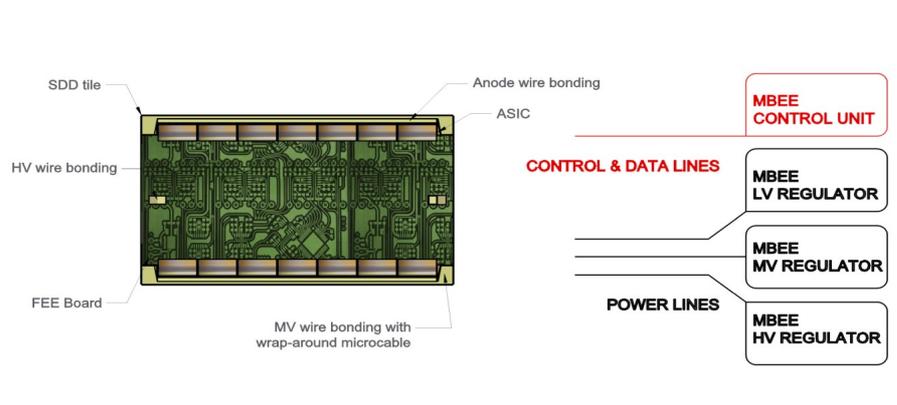

Figure 10. The back side of the LOFT detector, showing the front-end electronics and connections.

## 5. TELEMETRY AND INSTRUMENT MODES DESCRIPTION

The LAD scientific telemetry budget is estimated assuming default event-by-event data transmission, 24-bit per event. We conservatively assume a source with intensity 500 mCrab in the field of view at any time (this flux threshold includes >95% of the known X-ray sources with flux above 1 mCrab[12]. The expected count rate under this assumption is ~120,000 cts s$^{-1}$, in addition to the expected ~3000 cts s$^{-1}$ from the background.

Taking into account the typical net source exposure in LEO (~4000 s) and the full-orbit background counts, a total of ~11.5 Gbit are created over one orbit, corresponding to 1.9 Mbps orbit-average. This will be compressed to ~ 960 kbyte s$^{-1}$ through a lossless algorithm in the DHU. Preliminary simulations have been carried out using simulated LAD data streams and standard compression algorithms showing that a compression factor of ~2 is affordable. A dedicated study is ongoing within the team. As for the required computational resources, we foresee that a 64 Gbyte mass memory on the DHU will allow the temporary storage of excess telemetry.

*Orbital phases and Science Modes*

A typical orbit will consist of various orbital phases during which the LAD will operate in a default mode:

- It will be in "normal observing" mode, during the nominal orbit phase

- It will shift to the dedicated "SAA" mode, triggered by the Spacecraft itself, changing into a different setup during the higher radiation of the South Atlantic Geomagnetic Anomaly (SAA) when rates could be much higher from the background noise.

- It will use a Earth occultation mode, for cases in which the source is occulted by the Earth. The instrument will simply produce ratemeters and housekeeping or switch to a calibration mode to perform periodic electrical calibrations.

Some of the key science targets (~10 persistent sources and some bright X-ray transients) will have an average LAD count rate above $2 \times 10^5$ cts s$^{-1}$. Therefore we will employ a flexible set of data modes, as was done with the Event Data System (EDS) on RossiXTE. These science modes will allow the time and energy binning to be user-defined and optimized for the science goals within the available telemetry budget. The observing plan will be optimized by alternating bright and weak sources to allow for a gradual download of the excess telemetry, a strategy already successfully adopted by the RossiXTE satellite.

*Engineering modes*

In addition to the science modes, we foresee a series of operating engineering modes, as described in the following.

- An "OnTheFlyConfigure" mode in which it is possible to configure individual modules/detectors while simultaneously observing with others (triggers can be altered "on the fly").

- A "Diagnostic and Calibration" mode, that would send all information about events including absolute time-tags, raw charge data (as detected on anode before reconstruction), channel address. This mode will be activated for diagnostic and testing purposes as well as for module-by-module calibration.

- A "Pedestal" mode that would be aimed at measuring the mean value and rms noise of the electronics signal chains for each individual anode.

- An "Electrical Calibration" mode, composed of two sub-modes: the gain measurement and the threshold scan.

The strategy envisaged in the instrument baseline is based on a X-band telemetry down-link with a minimum net science data rate of 6.7Gbit/orbit. This is planned to be achieved by using the Kourou and Malindi ground stations.

## 6. FUTURE DIRECTIONS

The LOFT-LAD is undergoing continuous development as it moves toward the end of the assessment phase, and a number of different trade-offs are being studied and progressively frozen by the instrument team. The LOFT-LAD is a large and ambitious project, that requires the mass production of a large number of units, ultimately constituting a challenge in terms of procurement, assembly, implementation, verification (AIV) and calibration. These issues are being addressed early in the programme starting with the AIV philosophy and discussing the detailed manufacture plans with the equipment suppliers. The LOFT yellow book is currently required to be submitted to ESA by the end of 2012, and the results of the first down-selection of the M3 missions are expected by mid-2013. The down-selected candidates will then enter a definition phase (lasting ~1 year), at the end of which they will be subjected to a second, and final, down-selection which will identify the successful candidate for a launch in 2022.

The work of the MSSL and Leicester groups is supported by the UK Space Agency. The work of SRON is funded by the Dutch national science foundation (NWO). The work of the group at the University of Geneva is supported by the Swiss Space Office. The Italian team is grateful for support by ASI, INAF and INFN.The work of IAAT on LOFT is supported by the Germany's national research center for aeronautics and space DRL. The work of the IRAP group is supported by the French Space Agency. LOFT work at ECAP is supported by DLR under grant number 50 00 1111.